\documentclass[aps,pre,twocolumn,superscriptaddress]{revtex4-2} 

\usepackage{amsmath, amssymb}
\usepackage{xcolor}
\usepackage{graphicx} 
\usepackage{subcaption}
\usepackage[colorlinks=true, allcolors=blue]{hyperref}
\usepackage{indentfirst}

\begin{document}

\title{Synchronization and Swarming of Two-Mode Stochastic Oscillators}

\author{Szabolcs Vitus}
\affiliation{Department of Physics, Babeș-Bolyai University, M. Kogălniceanu street 1, 400084 Cluj-Napoca, Romania}

\author{Ferenc Járai-Szabó}
\email[]{ferenc.jarai@ubbcluj.ro}
\affiliation{Department of Physics, Babeș-Bolyai University, M. Kogălniceanu street 1, 400084 Cluj-Napoca, Romania}

\date{\today}

\begin{abstract}
Synchronization and swarming are canonical manifestations of self-organization, observable across scales from cellular processes to animal flocks. This study investigates the collective dynamics of a novel agent-based model where individuals exhibit both spatial mobility and internal, two-mode stochastic oscillatory states. By introducing a local, distance-dependent coupling between the agents' spatial configuration and their internal state transitions, we establish a mutual feedback loop that drives complex pattern formation. Through large-scale numerical simulations, we identify seven distinct morphological configurations, ranging from stationary \textit{Filled-disk} states to highly disordered \textit{Intense-motion} regimes. By performing a rigorous quantitative analysis of the rotational energy and radial dispersion, we transcend simple morphological classification and demonstrate that the system organizes into discrete, quantized topological attractors. We derive a macroscopic scaling law, $\Omega \propto r^{-1/2}$, which proves that the emerging rotating states are not rigid-body rotations, but rather composite differential vortex structures characterized by spontaneous chiral symmetry breaking. Our results suggest that these stable, quantized dynamical states are fundamental features of systems governed by bidirectional spatial-phase feedback, offering a robust framework for designing autonomous, decentralized robotic swarms.
\end{abstract}

\keywords{Swarmalators; Two-mode stochastic oscillators; Synchronization; Self-organization; Chiral symmetry breaking; Topological attractors}

\maketitle

\section{Introduction}

Self-organization in complex systems frequently manifests through two distinct but fundamentally related phenomena: synchronization in time and swarming in space. On one hand, synchronization governs processes ranging from the rhythmic flashing of fireflies \cite{buck1988} and the firing patterns of neural networks \cite{montbrio2015} to the spontaneous emergence of rhythmic applause \cite{neda2000}. Swarming, on the other hand, describes the collective, coordinated motion of self-propelled entities, such as flocks of birds \cite{bialek2012}, schools of fish, and marching locusts \cite{buhl2006}. Both phenomena rely heavily on the theoretical frameworks of statistical mechanics and nonlinear dynamics, focusing on how macroscopic order arises spontaneously from simple, localized interactions among a large number of agents.

Historically, these two domains have been investigated independently. Swarming models typically focus on the spatial trajectories and velocity alignment of agents, treating the entities as point particles without internal state dynamics \cite{vicsek1995,vicsek2012,marchetti2013}. Conversely, traditional synchronization models, most notably the Kuramoto paradigm \cite{kuramoto1975,acebron2005}, investigate the phase evolution of coupled oscillators while assuming fixed or irrelevant spatial topologies.

With the growing intersection of complex systems theory and decentralized robotics, several attempts have been made to unify these phenomena \cite{uriu2013,fujiwara2011,frasca2008}. In many of these mobile oscillator models, however, the coupling remains unidirectional: spatial proximity dictates interaction strength, but the phase dynamics do not feed back into the spatial motion. A significant breakthrough was the introduction of the "swarmalator" model by O'Keeffe et al. \cite{okeeffe2017}, which established a bidirectional feedback loop. In swarmalator systems, agents navigate their physical space while simultaneously updating their internal oscillatory phases; spatial attraction and repulsion are explicitly modulated by phase similarity, and phase synchronization is driven by spatial proximity. Since its inception, the swarmalator framework has inspired numerous extensions to capture more complex physical interactions, including the effects of time delay \cite{okeeffe2026delay}, inertia \cite{okeeffe2026inertia}, heterogeneous frequency-weighted couplings \cite{senthamizhan2026}, and directed, adaptive interactions \cite{yu2025}. 

In parallel, systems exhibiting discrete, mode-switching oscillatory dynamics have been modeled using two-mode stochastic oscillators \cite{neda2001}. These oscillators do not possess a continuously evolving natural frequency but instead alternate between two distinct states (e.g., a fast and a slow period) based on a global coupling parameter, successfully describing phenomena such as the transition to synchronized applause \cite{neda2000,neda2001}.

In this paper, we propose a novel agent-based model that bridges these two frameworks by investigating the collective dynamics of two-mode stochastic oscillators endowed with swarmalator-like spatial mobility. We replace the conventional global coupling of two-mode oscillators with a local, distance-dependent interaction field. This establishes an intricate feedback mechanism. The spatial arrangement of the agents dictates the local pulse intensity they perceive, which directly triggers their state transitions. Concurrently, their discrete mode-switching alters their effective phase, subsequently driving their spatial aggregation and motion. 

While continuous-phase swarmalator models successfully capture smooth spatial-phase synchronization, they inherently assume uninterrupted, gradient-based information transfer between agents. However, communication in many biological and physical systems relies on discrete pulses subject to local stochastic fluctuations \cite{mirollo1990}. Prominent examples include flashing fireflies, firing neurons, and threshold-based chemical oscillators. Recognizing this limitation, very recent work has begun to explore pulsatile interactions and Winfree-type couplings within the swarmalator paradigm \cite{yadav2025collective,Ghosh2025}. Building on this crucial shift toward discrete communication, we take a step further: by replacing the continuous Kuramoto-like phase evolution entirely with discrete, state-switching stochastic oscillators governed by a local density threshold ($f^*$), our model captures the highly non-linear, integrate-and-fire nature of these real-world active systems.

Through large-scale numerical simulations, we systematically map the parameter space of this system and reveal a rich variety of emergent macroscopic states. Notably, we demonstrate that the system spontaneously breaks chiral symmetry, organizing into quantized topological attractors characterized by differential rotation and specific scaling laws.

\section{The Model and Computational Details}
\label{sec:model}

To investigate the interplay between spatial swarming and internal phase synchronization, we construct a unified framework that combines the distance-dependent spatial dynamics of swarmalators with the discrete, state-switching internal dynamics of two-mode stochastic oscillators. 

\subsection{The Two-Mode Swarmalator Model}

We consider a two-dimensional system of $N$ identical agents, where each agent $i$ ($i = 1, \dots, N$) is characterized by its spatial position vector $\mathbf{x}_i \in \mathbb{R}^2$ and an internal phase $\theta_i \in [0, 2\pi)$. The spatial evolution of the agents is governed by an overdamped equation of motion, adopted from the original swarmalator model \cite{okeeffe2017}. It is driven by a combination of phase-dependent attraction and phase-independent short-range repulsion:
\begin{equation}
    \dot{\mathbf{x}}_i = \frac{1}{N} \sum_{j \neq i}^{N} \left[ \frac{\mathbf{x}_j - \mathbf{x}_i}{|\mathbf{x}_j - \mathbf{x}_i|} \big(1 + J \cos(\theta_j - \theta_i)\big) - \frac{\mathbf{x}_j - \mathbf{x}_i}{|\mathbf{x}_j - \mathbf{x}_i|^2} \right].
    \label{eq:spatial}
\end{equation}
The first term in the brackets represents a normalized attractive force, the strength of which is modulated by the phase similarity of the interacting agents through the coupling parameter $J$. For $J > 0$, agents with similar phases attract each other more strongly. The second term ensures structural stability by providing a strong, short-range repulsive force that prevents collisions and prevents the distance $|\mathbf{x}_j - \mathbf{x}_i|$ from collapsing to zero.

It should be noted that the spatial equation of motion is strictly deterministic and omits explicit spatial fluctuations, such as an additive Langevin noise term. This approach maintains consistency with the foundational swarmalator framework \cite{okeeffe2017}. From a physical perspective, exogenous spatial noise is omitted because the system is already subjected to strong endogenous fluctuations. As will be detailed below, the internal phase transitions are highly stochastic. Due to the phase-dependent coupling in Eq.~\eqref{eq:spatial}, these random internal state jumps immediately propagate into the spatial domain, manifesting as abrupt fluctuations in the inter-agent attractive forces.

Unlike classical continuous phase oscillators, the internal dynamics of our agents are modeled as two-mode stochastic oscillators \cite{neda2001}. Each oscillator progresses through a cycle consisting of three consecutive states: $A \to B \to C \to A$. The instantaneous phase is defined as $\theta_i(t) = 2\pi (t_p / T_i)$, where $t_p$ is the proper time elapsed since the start of the current cycle, and $T_i$ is the total period of the selected mode.

The duration of state A, denoted as $\tau_A$, represents the stochastic component of the cycle and is drawn from an exponential probability distribution $P(\tau_A) = (1/\tau^*) \exp(-\tau_A / \tau^*)$, where $\tau^*$ is the characteristic time. State B is a waiting period, the duration of which defines the operational mode of the oscillator. The oscillator can operate in either Mode 1 (short waiting time, $\tau_{B1}$) or Mode 2 (long waiting time, $\tau_{B2} > \tau_{B1}$). Finally, during state C (duration $\tau_C$), the oscillator emits a discrete pulse. The pulse intensity $f_j(t)$ of the $j$-th oscillator is $1/N$ during state C, and $0$ otherwise.

The fundamental novelty of our model lies in the local, distance-dependent feedback mechanism that dictates the mode-switching. The effective pulse intensity $p_i(t)$ perceived by the $i$-th oscillator is weighted by the inverse of the Euclidean distance to all other emitting oscillators:
\begin{equation}
    p_i(t) = \sum_{j \neq i}^{N} \frac{f_j(t)}{|\mathbf{x}_j - \mathbf{x}_i|}.
    \label{eq:pulse}
\end{equation}
At the exact moment state A concludes, the $i$-th oscillator evaluates this local field against a predefined global threshold parameter, $f^*$. If $p_i < f^*$, the agent transitions into Mode 1 (attempting to increase the local pulse density by adopting a shorter period). Conversely, if $p_i \geq f^*$, it transitions into Mode 2 (adopting a longer period). This creates a bidirectional feedback loop: the agents' spatial arrangement determines their perceived pulse intensity and internal mode, while their internal mode dictates their phase, which in turn drives their spatial aggregation via Eq.~\eqref{eq:spatial}.

\subsection{Numerical Integration and Simulation Protocol}

The system features continuous spatial trajectories governed by ordinary differential equations alongside discrete, stochastic jumps in the phase dynamics. This complex hybrid nature requires a robust numerical approach.

We simulated a system of $N = 1000$ agents. To preserve the mathematical integrity of the numerical integration across non-differentiable events, we employed an operator splitting approach. During each time step of $dt = 0.1$, the spatial equations of motion \eqref{eq:spatial} were advanced using a fourth-order Runge-Kutta (RK4) scheme while the internal phases and modes of the agents were strictly held constant. This ensures the spatial ODE remains smooth during the intermediate RK4 evaluations. The discrete stochastic jumps, mode transitions, and phase updates were subsequently evaluated and executed only after the spatial integration step was completed. The step size $dt$ was chosen to be sufficiently small to resolve the coupled dynamics accurately while ensuring computational efficiency for large-scale parameter sweeps. The temporal parameters of the oscillators were fixed across all simulations to $\tau^* = 1.0$, $\tau_{B1} = 4.0$, $\tau_{B2} = 8.0$, and $\tau_C = 1.0$.

For each simulation run, the system was initialized with random uniform spatial coordinates within a bounding domain of $\mathbf{x}_i \in [-1.5, 1.5] \times [-1.5, 1.5]$ to establish a controlled initial density. However, the subsequent spatial evolution of the agents was simulated using open (infinite) boundary conditions. Confinement within a finite simulation box was deliberately avoided to ensure that the emergent morphological patterns were not artifacts of spatial constraints. Instead, the structural cohesion of the swarm is maintained entirely by the self-organizing balance between the phase-dependent attractive and short-range repulsive interactions defined in Eq.~\eqref{eq:spatial}.

The parameter space defined by the spatial coupling strength $J \in [-1.0, 1.0]$ and the target intensity threshold $f^* \in [0.0, 0.3]$ was systematically mapped. To verify the stability of the topological states against initial random conditions, an ensemble of three independent simulation runs was performed for each $(J, f^*)$ parameter pair. While a small ensemble size precludes the precise characterization of diverging fluctuations exactly at the phase boundaries, it is highly sufficient for our objective of mapping the stable attractors. This is because the macroscopic observables investigated are subjected to extensive self-averaging: they are computed over a large system size and strictly time-averaged over long temporal windows after the system has reached a stationary dynamical state.

\section{Phase Space and Synchronization}
\label{sec:phase_space}

To understand the macroscopic behavior emerging from the microscopic feedback loop, we systematically explored the parameter space defined by the phase-dependent spatial coupling strength $J \in [-1.0, 1.0]$ and the target global pulse intensity $f^* \in [0.0, 0.3]$. 

\begin{figure*}[htpb]
    \centering
    \includegraphics[width=0.9\textwidth]{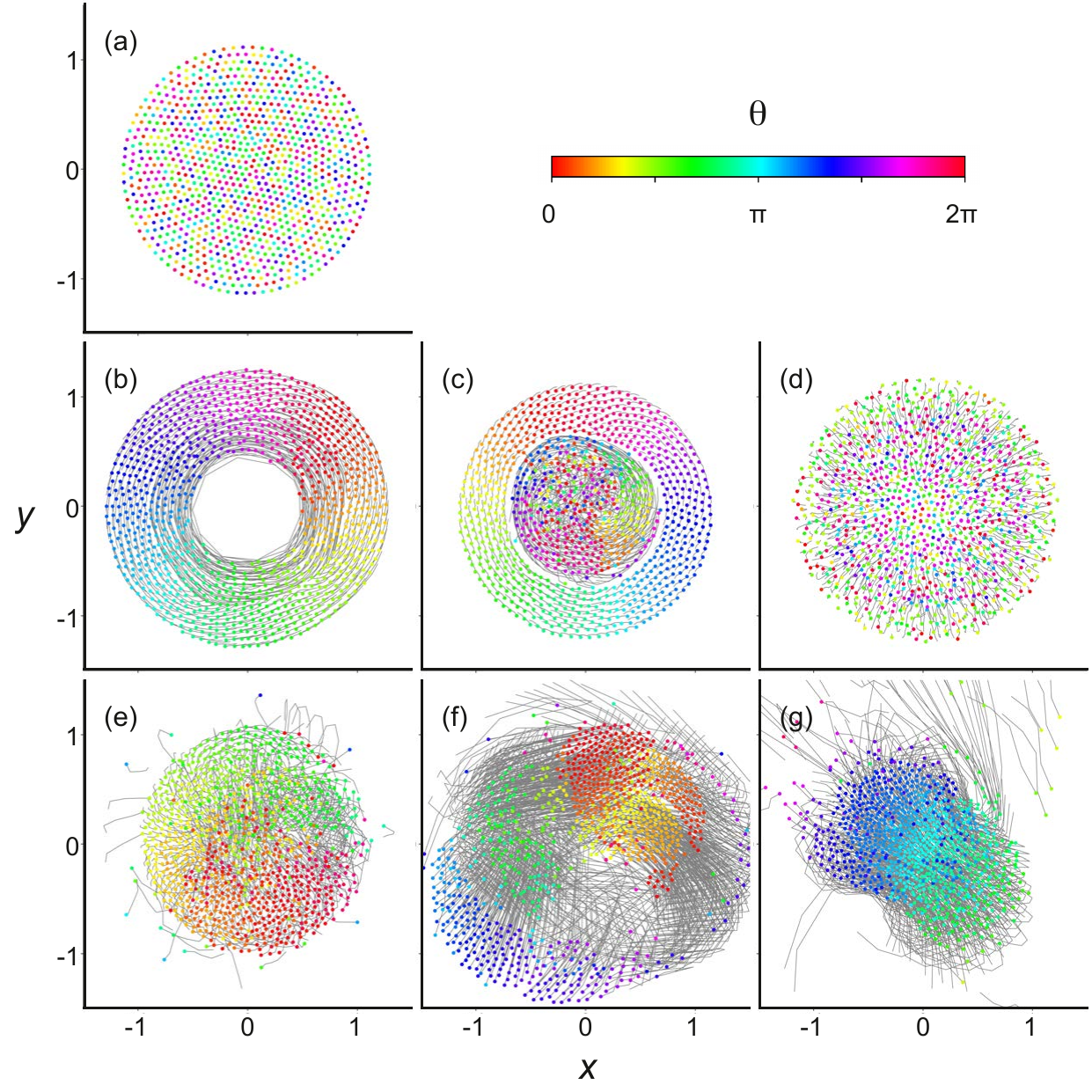}
    \caption{Representative snapshots of the distinct macroscopic morphological states emerging in the two-mode swarmalator system. The color scale indicates the instantaneous internal phase of the oscillators, highlighting the resulting spatial phase segregation of the agents. The accompanying motion trails depict the short-time trajectories of the oscillators, effectively capturing the local kinematic dynamics and macroscopic rotation. The identified states are: (a) the stationary \textit{Filled-disk}, (b) the chiral \textit{Hollow-ring}, (c) the chiral \textit{Filled-ring}, (d) the radially oscillating \textit{Breathing} state, (e) the \textit{Mild-motion} state, (f) the highly disordered \textit{Intense-motion} state, and (g) the anisotropic \textit{Pulsating-spindle} state. Videos capturing the real-time dynamics of these morphological states are available in the Supplemental Material \cite{supp_mat}.}
    \label{fig:morphology_snapshots}
\end{figure*}

Initial qualitative observations of the agents' spatial arrangements and mobility revealed seven morphologically distinct dynamical regimes, as illustrated in Fig.~\ref{fig:morphology_snapshots}. In the non-rotating domain ($J \leq 0$), the system predominantly forms a static, homogeneously distributed \textit{Filled-disk}, or a radially oscillating \textit{Breathing} state. In the domain characterized by strong phase-spatial feedback ($J > 0$), the system exhibits spontaneous chiral symmetry breaking, organizing into rotating topological configurations. These include the \textit{Filled-ring} and \textit{Hollow-ring} states, as well as highly disordered, active regimes such as \textit{Mild-motion} and \textit{Intense-motion}. Furthermore, at high spatial coupling, the system can break radial symmetry entirely, forming an elongated, anisotropic \textit{Pulsating-spindle} state.

To quantitatively characterize the temporal coherence of the system across these diverse morphological states, we examine the global pulse intensity, $f(t) = \sum_{i=1}^{N} f_i(t)$. If the agents are globally synchronized, $f(t)$ exhibits a strong periodic macroscopic signal; otherwise, the signal resembles asynchronous noise. We quantify the degree of self-similarity and periodicity in the collective signal by introducing a normalized time-delayed prediction error function, $\Delta(T)$:
\begin{equation}
    \Delta(T) = \frac{1}{2M} \lim_{x \to \infty} \frac{1}{x} \int_{0}^{x} |f(t) - f(t+T)| dt,
    \label{eq:error_function}
\end{equation}
where the normalization factor $M$ is the mean absolute deviation of the signal from its temporal average $\langle f(t) \rangle$:
\begin{equation}
    M = \lim_{x \to \infty} \frac{1}{x} \int_{0}^{x} |f(t) - \langle f(t) \rangle| dt.
    \label{eq:normalization}
\end{equation}
The global minimum of the error function, $\Delta_m = \min[\Delta(T)]$, evaluated at the optimal delay $T_m$ (which corresponds to the macroscopic period of the system), provides a robust inverse measure of temporal coherence. We thus define the periodicity measure $p$ as:
\begin{equation}
    p = \frac{1}{\Delta_m}.
    \label{eq:periodicity}
\end{equation}
The parameter $p$ takes values in the interval $[1, \infty)$. A value of $p \approx 1$ indicates a complete lack of temporal correlation (an asynchronous state), whereas $p \gg 1$ denotes a highly synchronized regime with strong temporal ordering of the phases.

\begin{figure*}[htpb]
    \centering
    \begin{minipage}[t]{0.48\textwidth}
        \vspace{0pt}
        \subcaption{}
        \includegraphics[width=\textwidth]{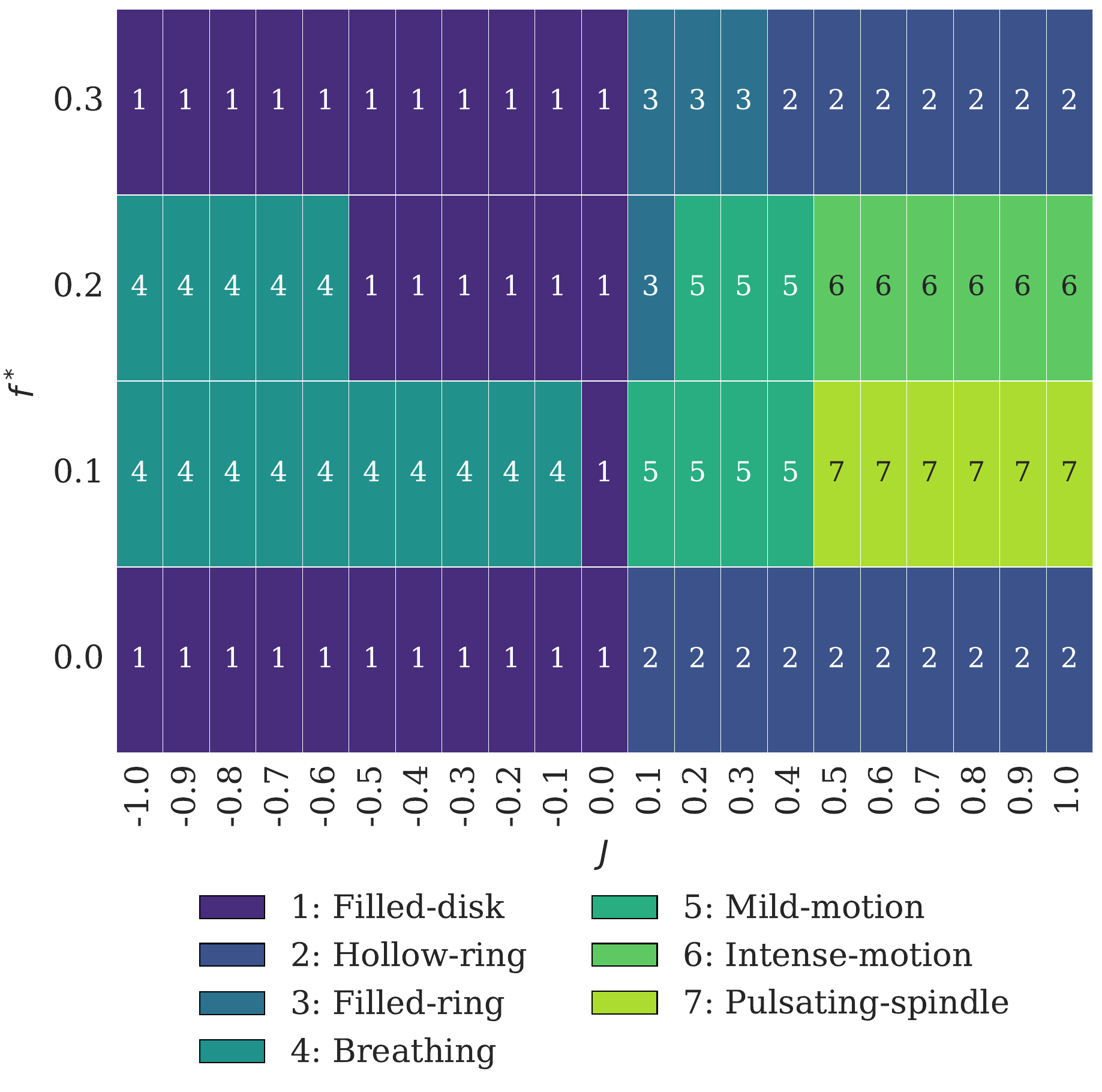}
    \end{minipage}
    \hfill
    \begin{minipage}[t]{0.49\textwidth}
        \vspace{0pt}
        \subcaption{}
        \includegraphics[width=\textwidth]{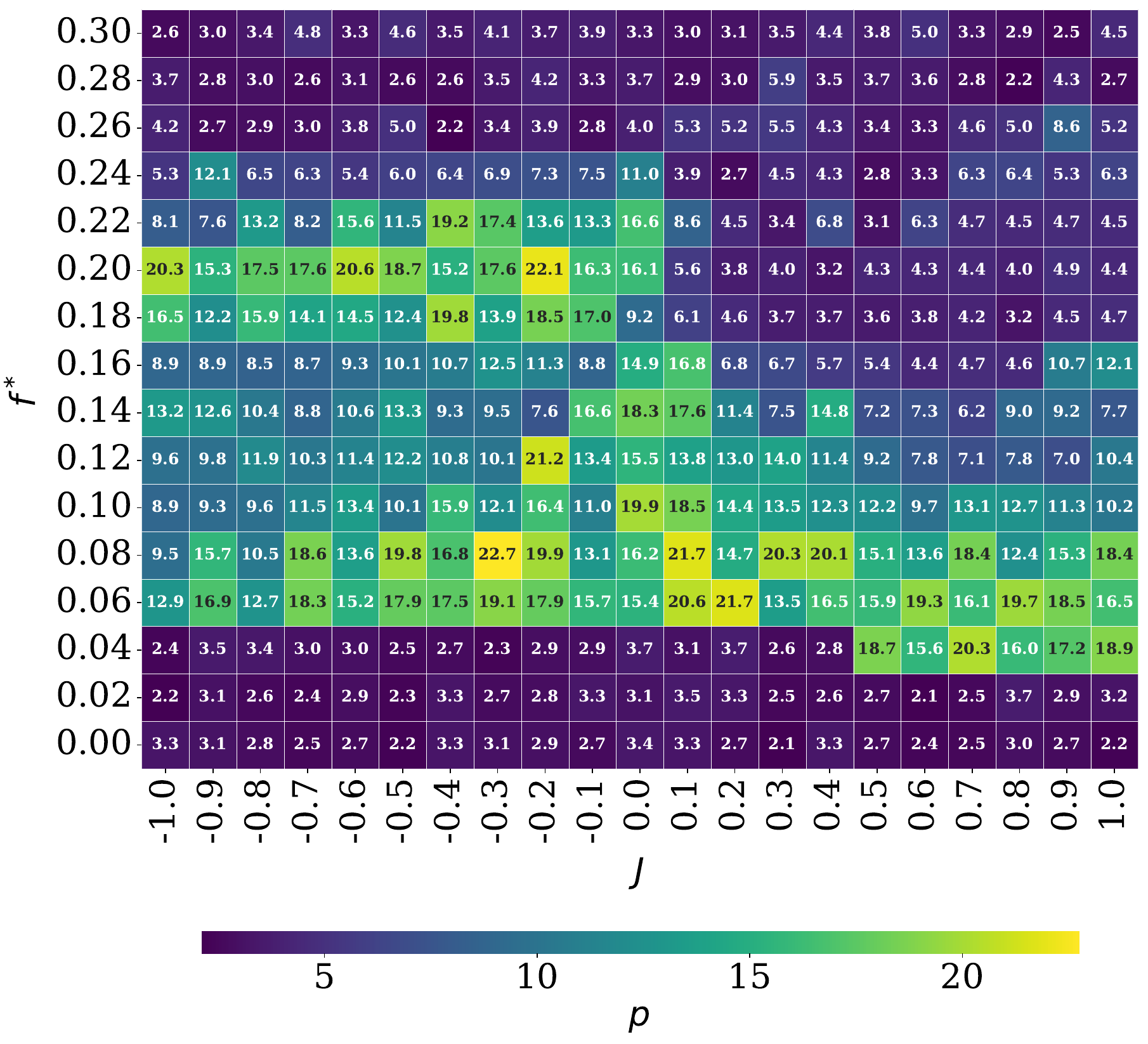}
    \end{minipage}
    \caption{Macroscopic states of the two-mode swarmalator system in the $(J, f^*)$ parameter space for $N=1000$ agents. (a) Qualitative phase diagram indicating the seven identified morphological regimes. (b) Synchronization phase diagram showing the periodicity measure $p$. A sharp transition from synchronized ($p \gg 1$, lighter colors) to asynchronous ($p \approx 1$, dark colors) dynamics is observed, particularly where spatial swarming severely interferes with phase alignment.}
    \label{fig:phase_diagrams}
\end{figure*}

The resulting phase diagrams, detailing both the morphological states and the temporal synchronization measure $p$, are presented in Fig.~\ref{fig:phase_diagrams}. The synchronization map reveals a pronounced separation between the highly ordered and asynchronous domains. In the baseline two-mode oscillator model lacking spatial dynamics, global coherence is generally maintained. However, in our unified model, we observe a distinct breakdown of synchronization for parameter values around $f^* \approx 0.2$ specifically when $J > 0$. 

This localized loss of synchronization represents a fundamental departure from continuous-phase swarmalator models, where spatial aggregation typically stabilizes global phase coherence. Because our agents operate via a rigid threshold mechanism rather than a continuous gradient, the formation of dense, co-rotating clusters induces violent, highly localized fluctuations in the perceived pulse density $p_i(t)$. These stochastic threshold crossings dynamically frustrate global phase alignment. Consequently, spatial swarming acts as an active symmetry-breaking mechanism that destroys temporal coherence, a phenomenon uniquely captured by the discrete mode-switching feedback loop.

\section{Vortex States and Differential Rotation}
\label{sec:vortex_states}

To uncover the physical mechanisms governing the chiral states in the $J > 0$ regime, we perform a quantitative analysis of the macroscopic spatial and kinematic properties of the swarm. The most informative characterization of the collective rotational modes is obtained by investigating the relationship between the time-averaged macroscopic rotational energy, $\langle E_{\varphi} \rangle_t$, and the time-averaged radial dispersion, $\langle \sigma_r \rangle_t$. 

We define the specific macroscopic rotational energy (energy per oscillator) arising from the tangential velocities as:
\begin{equation}
    E_{\varphi} = \frac{1}{N} \sum_{i=1}^{N} v_{\varphi, i}^2 = \langle v_{\varphi}^2 \rangle,
    \label{eq:rot_energy}
\end{equation}
and the radial dispersion (spatial variance) as:
\begin{equation}
    \sigma_r = \sqrt{ \frac{1}{N} \sum_{i=1}^{N} (r_i - \langle r \rangle)^2 }.
    \label{eq:radial_disp}
\end{equation}

To establish a quantitative basis for comparing different dynamical states, we first normalize the relevant time-averaged macroscopic observables to the interval $[0, 1]$ using min-max scaling. 

An analysis of the distribution of the time-averaged normalized radial dispersion, $\langle \sigma_r \rangle_{norm}$, reveals a prominent peak at $\langle \sigma_r \rangle^*_{norm} \approx 0.2087$. Physically, this finite minimum dispersion arises from the strong short-range repulsion term in Eq.~\eqref{eq:spatial}. This repulsion enforces an excluded volume and prevents the swarm from collapsing into a single point. This value corresponds to the spatial variance of the non-rotating ground state, identified as the \textit{Filled-disk} configuration. 

When phase-dependent spatial coupling initiates macroscopic rotation, the swarm undergoes centrifugal expansion. To isolate the spatial variance driven exclusively by this rotational dynamics, we define the excess radial dispersion, $\tilde{\sigma}_r$, by shifting the distribution such that the non-rotating ground state is relocated to the origin:
\begin{equation}
    \tilde{\sigma}_r = \langle \sigma_r \rangle_{norm} - \langle \sigma_r \rangle^*_{norm}.
    \label{eq:shifted_dispersion}
\end{equation}

With the transformed macroscopic variables defined, we establish a theoretical baseline by formulating the null hypothesis of macroscopic rigid-body rotation. If the swarm were to rotate as a rigid body, all agents would revolve around the center of mass with an identical and constant angular velocity $\Omega$. The tangential velocity of the $i$-th oscillator would be $v_{\varphi, i} = \Omega r_i$. Substituting this into Eq.~\eqref{eq:rot_energy} yields $E_{\varphi} = \Omega^2 \langle r^2 \rangle$. Using the definition of variance, $\langle r^2 \rangle = \sigma_r^2 + \langle r \rangle^2$, the rotational energy under the rigid-body assumption becomes:
\begin{equation}
    E_{\varphi} = \Omega^2 (\sigma_r^2 + \langle r \rangle^2).
    \label{eq:rigid_body}
\end{equation}
This derivation indicates that for rigid-body rotation, the macroscopic rotational energy scales quadratically with the excess radial dispersion ($E_{\varphi} \propto \tilde{\sigma}_r^2$). 

\begin{figure}[htpb]
    \centering
    \includegraphics[width=0.5\textwidth]{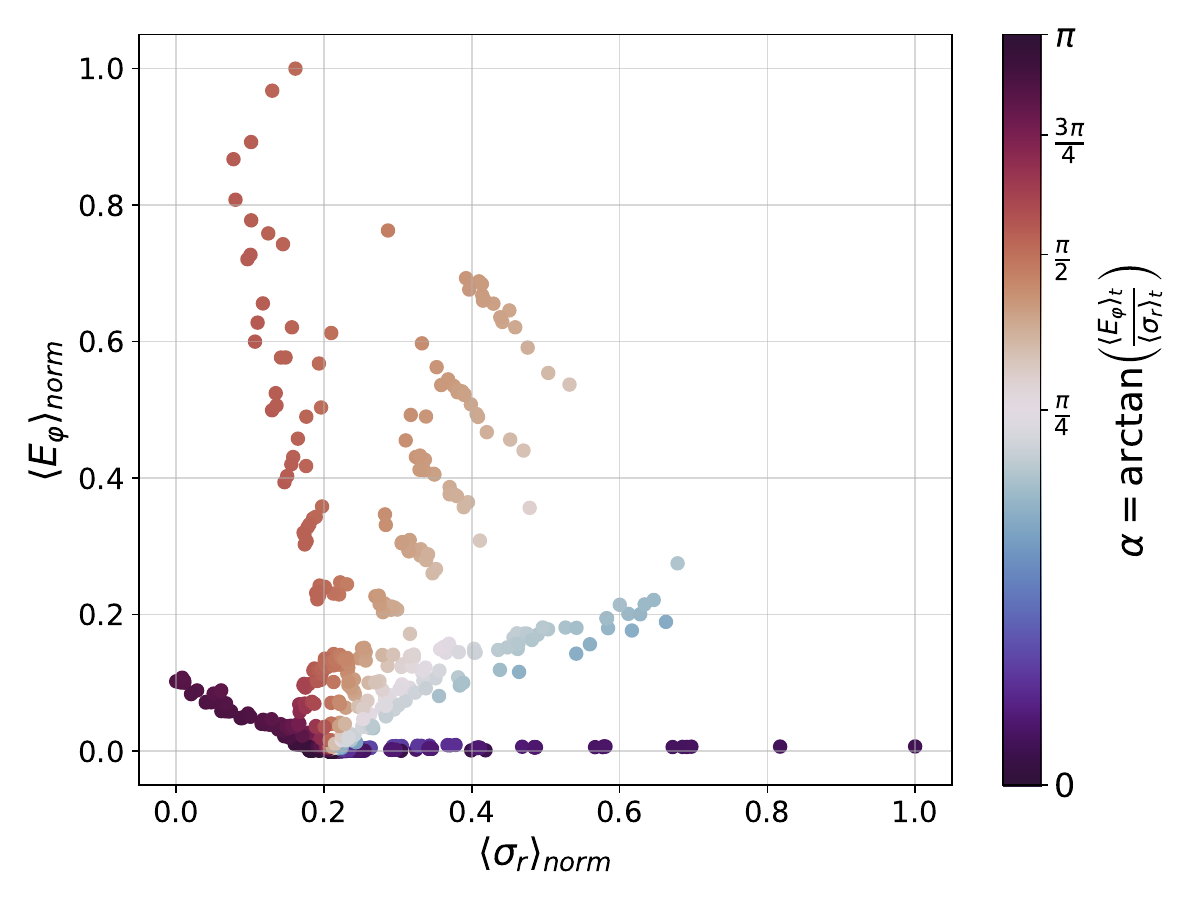}
    \caption{Scaled and time-averaged rotational energy $\langle E_{\varphi} \rangle_{norm}$ versus the time-averaged normalized radial dispersion $\langle \sigma_r \rangle_{norm}$. Each point represents a distinct simulation run. The plot includes three independent realizations for every $(J, f^*)$ parameter pair across the entire explored parameter space $J \in [-1.0, 1.0]$. The data points collapse onto highly linear radial structures intersecting at $\langle \sigma_r \rangle^*_{norm}$, indicating discrete topological vortex states with quantized vortex strengths. Colors map to the chiral order parameter $\alpha$.}
    \label{fig:scatter_energy}
\end{figure}

However, our extensive numerical simulations reveal a fundamentally different behavior. Figure~\ref{fig:scatter_energy} displays the time-averaged rotational energy against the excess radial dispersion for all independent simulation runs across the entire parameter space. The data points collapse onto highly linear radial structures that intersect at the origin. This empirical observation dictates a purely linear scaling law:
\begin{equation}
    E_{\varphi} \propto \tilde{\sigma}_r.
    \label{eq:empirical_scaling}
\end{equation}
This striking linear correlation unequivocally disproves the rigid-body rotation hypothesis. 

To explain the linear scaling, we propose that the swarm undergoes differential rotation and organizes into topological vortex states. We hypothesize an angular velocity profile that decays with the square root of the distance from the center, $\Omega(r) \propto r^{-1/2}$. For an oscillator at radius $r_i$, the angular velocity is $\Omega_i = k \cdot r_i^{-1/2}$, where $k$ is a constant denoting the vortex strength. The tangential velocity is then:
\begin{equation}
    v_{\varphi, i} = \Omega_i r_i = \left( k \cdot r_i^{-1/2} \right) r_i = k \cdot r_i^{1/2}.
    \label{eq:vortex_vel}
\end{equation}
Substituting Eq.~\eqref{eq:vortex_vel} into the macroscopic rotational energy Eq.~\eqref{eq:rot_energy}, we obtain:
\begin{equation}
    E_{\varphi} = \frac{1}{N} \sum_{i=1}^{N} (k \cdot r_i^{1/2})^2 = k^2 \frac{1}{N} \sum_{i=1}^{N} r_i = k^2 \langle r \rangle.
    \label{eq:vortex_energy}
\end{equation}
Assuming geometric self-similarity during the centrifugal expansion of a given dynamical phase, the excess macroscopic radius $\Delta \langle r \rangle$ is strictly proportional to the excess radial dispersion $\tilde{\sigma}_r$. Therefore, the rotational kinetic energy scaling reduces to:
\begin{equation}
    E_{\varphi} \propto k^2 \tilde{\sigma}_r.
    \label{eq:theoretical_scaling}
\end{equation}

Equation \eqref{eq:theoretical_scaling} is in perfect agreement with the empirically observed linear scaling \eqref{eq:empirical_scaling}. It proves that the rotating swarmalator states are composite topological vortices characterized by a fast-rotating dense core and a progressively slower-rotating outer shell. 

Furthermore, the linear structures observed in Fig.~\ref{fig:scatter_energy} are highly discrete. We define a chiral order parameter $\alpha$ as the directional tangent of these linear structures:
\begin{equation}
    \alpha = \arctan \left( \frac{\langle E_{\varphi} \rangle_{norm}}{\tilde{\sigma}_r} \right).
    \label{eq:alpha}
\end{equation}
The discrete nature of the angles $\alpha$ confirms that the vortex strengths ($k^2$) are quantized. The system cannot maintain an arbitrary rotational topology. Driven by the bidirectional feedback loop, the swarm inevitably collapses into one of several distinct topological attractors.

To rigorously verify the quantization of the vortex states, we examine the statistical distribution of the chiral order parameter $\alpha$ across the entire explored parameter space. 

\begin{figure*}[htpb]
    \centering
    \begin{minipage}[t]{0.5\textwidth}
        \vspace{0pt} 
        \subcaption{} 
        \includegraphics[width=\textwidth]{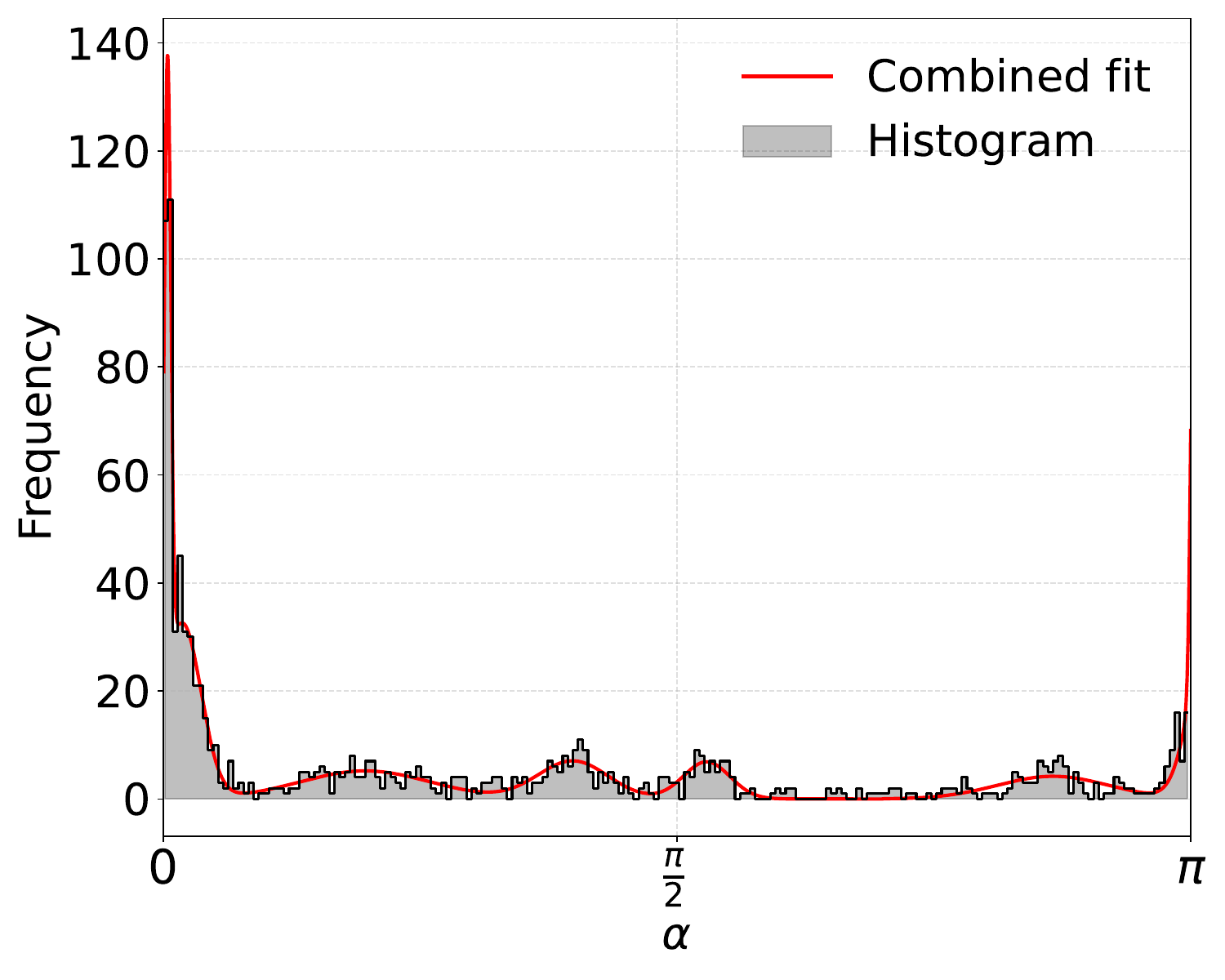}
    \end{minipage}
    \hfill
    \begin{minipage}[t]{0.48\textwidth}
        \vspace{0pt} 
        \subcaption{} 
        \includegraphics[width=\textwidth]{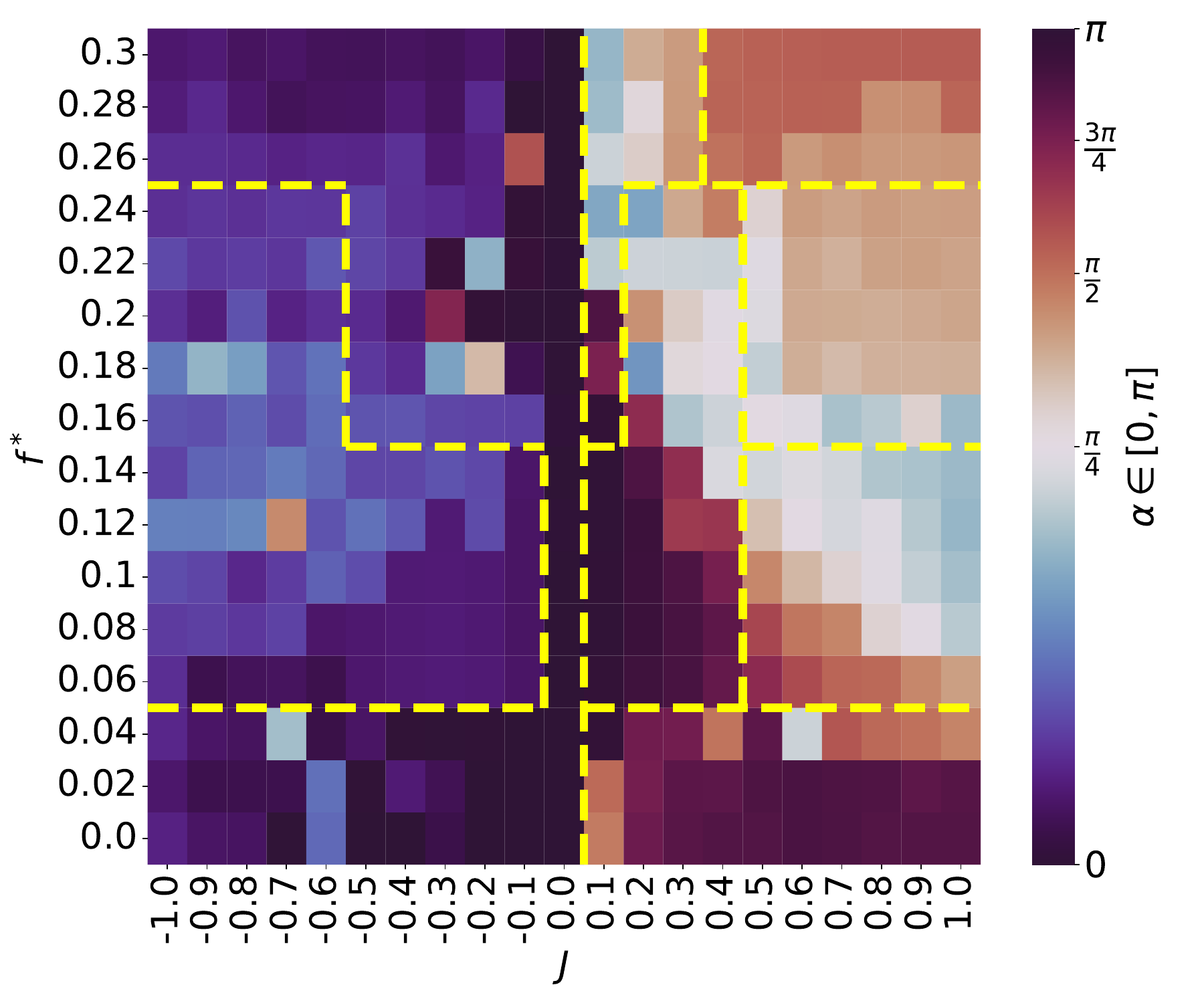}
    \end{minipage}
    \caption{Quantization of topological attractors and the $\alpha$ phase diagram. (a) Angular distribution of $\alpha$ fitted with a sum of Gaussian components. The discrete peaks confirm that the system relaxes into quantized vortex states. (b) Phase diagram showing the values of $\alpha$ in the $(J, f^*)$ parameter space, utilizing a power-law color normalization ($\gamma = 0.5$) to enhance contrast for $J<0$. Sharp boundaries indicate transitions between different topological attractors. The dashed lines indicate the boundaries of the dynamical states identified via prior qualitative analysis.}
    \label{fig:alpha_analysis}
\end{figure*}

As shown in Fig.~\ref{fig:alpha_analysis}(a), the probability density of $\alpha$ does not form a continuous, smeared spectrum. Instead, it decomposes into well-separated, discrete components that can be accurately fitted by a sum of Gaussian curves. The dominant peak at $\alpha \approx 0$ corresponds to the non-rotating states (such as the \textit{Filled-disk} and \textit{Breathing} configurations), where the macroscopic rotational energy is negligible. The remaining distinct peaks represent the stable, rotating topological attractors. This discrete spectrum provides definitive proof that the chiral symmetry breaking in the swarmalator system is quantized: the swarm cannot sustain arbitrary differential rotation but spontaneously relaxes into specific, stable composite vortex states.

To understand the boundaries and stability of these attractors, we map the time-averaged $\alpha$ values back onto the macroscopic control parameters, yielding the phase diagram presented in Fig.~\ref{fig:alpha_analysis}(b). The most prominent feature is the severe discontinuity at $J = 0$. For $J \leq 0$, the negative phase-dependent spatial coupling suppresses coherent rotating waves, trapping the system in static or purely radially oscillating states ($\alpha \approx 0$). 

In the active chiral regime ($J > 0$), the parameter $J$ drives the rotation, while the target pulse intensity $f^*$ acts as a fine-tuning control that forces the system through discrete phase transitions. The diagram reveals sharply defined, island-like basins of attraction. The abrupt changes in the $\alpha$ value across these boundaries are classic signatures of phase transitions in statistical physical systems, where microscopic feedback rules enforce macroscopically stable, discrete configurations. It should be noted that because all simulations were initialized from random macroscopic states at a fixed system size ($N = 1000$), we did not perform the adiabatic parameter sweeps or finite-size scaling necessary to classify the strict order (e.g., first-order versus continuous) of these non-equilibrium transitions. The precise characterization of the critical fluctuations and potential hysteresis loops at these boundaries remains an important avenue for future investigation. 

Crucially, the $\alpha$ phase diagram exposes hidden topological differences that qualitative visual analysis cannot resolve. For instance, visually similar \textit{Hollow-ring} configurations emerge in disparate regions of the right-hand parameter space. However, their distinct $\alpha$ values in Fig.~\ref{fig:alpha_analysis}(b) demonstrate that these geometrically similar structures belong to fundamentally different dynamical attractors, characterized by varying vortex strengths and internal shear.

\section{Conclusions and Perspectives}
\label{sec:conclusions}

In this work, we introduced a unified agent-based model that establishes a bidirectional feedback loop between the spatial dynamics of swarmalators and the internal state transitions of two-mode stochastic oscillators. By systematically mapping the parameter space, we uncovered a rich spectrum of collective behaviors. Our approach demonstrates that complex population-level movement patterns do not require distinct behavioral rules for every specific motion type. Instead, diverse macroscopic patterns can emerge spontaneously from the robust interplay between spatial organization and internal phase dynamics.

The most significant physical finding of this study is the spontaneous breaking of chiral symmetry in the presence of positive phase-dependent spatial coupling, a phenomenon of broad interest in the study of chiral active matter \cite{lowen2016}. Our quantitative analysis unequivocally disproves the assumption of macroscopic rigid-body rotation for the swarming agents. We derived and empirically validated through simulations a macroscopic scaling law indicating that the swarm undergoes differential rotation. The system organizes into composite topological vortex states where the angular velocity decays with the distance from the center according to the relation $\Omega \propto r^{-1/2}$.

Furthermore, by introducing the chiral order parameter $\alpha$, we demonstrated that the rotational modes of the system are heavily quantized. The parameter space does not support a continuous spectrum of angular velocities. Driven by the local feedback mechanism, the swarm inevitably relaxes into one of several discrete, stable topological attractors. This diagnostic parameter also revealed hidden dynamical complexities that remain invisible to purely morphological observations. Geometrically similar configurations frequently possess fundamentally different topological characteristics and are distinguished by distinct vortex strengths and internal shear distributions.

Crucially, this quantization of vortex strengths highlights the unique predictive power of the two-mode stochastic framework. In models utilizing continuous phase dynamics, macroscopic rotational modes often span a continuous spectrum dictated by initial conditions and coupling strengths. In contrast, the binary restriction of the agents' internal operational modes combined with the rigid transition threshold effectively discretizes the allowed macroscopic shear forces. The bidirectional feedback loop thereby forces the swarm to collapse exclusively into one of several topologically quantized attractors.

From an applied perspective, the principles of decentralized self-organization elucidated in this model hold substantial promise for the field of robotics. The ability to coordinate complex, stable, and adaptive flow patterns without a central control unit is a critical requirement for the deployment of autonomous robotic swarms. In particular, systems governed by such simple local interaction rules offer highly viable frameworks for nanorobotics. In these environments, physical size constraints demand strictly decentralized control strategies for critical tasks such as targeted drug delivery within the human body.

\section*{Data Availability}
The datasets generated during this study, the associated C++ simulation code, and the supplemental videos of the dynamical states are openly available in the Zenodo repository at \url{https://doi.org/10.5281/zenodo.21075696}.

\bibliography{ws-references} 
\end{document}